\documentstyle[twocolumn,epsf,seceq]{jpsj}

\def\nle{\ \raise.3ex\hbox{$<$}\kern-0.8em\lower.7ex\hbox{$\sim$}\ }
\def\nge{\ \raise.3ex\hbox{$>$}\kern-0.8em\lower.7ex\hbox{$\sim$}\ }

\def\chione{\chi'(\omega; t)}
\def\chitilde{\tilde{\chi}(\omega; t)} 
\def\chitwo{\chi''(\omega; t)}
\def\chitwoeq{\chi''_{\rm eq}(\omega)}
\def\Lomega{L_T(\tomega)}

\def\lovlp{l_{\Delta T}}
\def\Tc{T_{\rm c}}

\def\tshone{t_{\rm sh1}}
\def\tshtwo{t_{\rm sh2}}
\def\tomega{t_{\omega}}
\def\tw{t_{\rm w}}
\def\twone{t_{\rm w1}}

\def\twoneeff{t_{\rm w1}^{\rm eff}}
\def\twtwo{t_{\rm w2}}
\def\twtwoeff{t_{\rm w2}^{\rm eff}}

\def\runtitle{Numerical Study on Aging Dynamics in  the 3D Ising
  Spin-Glass Model. III.}
\def\runauthor{Hajime {\sc Takayama} and Koji {\sc Hukushima} }

\title{Numerical Study on Aging Dynamics in  the 3D Ising Spin-Glass
   Model.\\
   III. Cumulative Memory and `Chaos' Effects \\ 
   in the Temperature-Shift Protocol}
\author{
Hajime {\sc Takayama}\footnote{E-mail:
takayama@issp.u-tokyo.ac.jp} and Koji {\sc
Hukushima}\footnote{E-mail: 
hukusima@issp.u-tokyo.ac.jp}}
\inst{Institute for Solid State Physics, University of Tokyo,
5-1-5 Kashiwa-no-ha, Kashiwa, Chiba 277-8581\ \ }
\recdate{\today}
\abst{
The temperature ($T$)-shift protcol of aging in the 3 dimensional (3D)
Edwards-Anderson (EA) spin-glass  
(SG) model is studied through the out-of-phase component of the ac 
susceptibility simulated by the Monte Carlo method. For processes with a 
small magnitude of the $T$-shift, $\Delta T$, the memory imprinted 
before the $T$-shift is preserved under the $T$-change and the SG 
short-range order continuously grows after the $T$-shift, which we call 
the cumulative memory scenario. For a negative $T$-shift process with a 
large $\Delta T$ the deviation from the cumulative memory scenario has 
been observed for the first time in the numerical simulation. We 
attribute the phenomenon to the `chaos effect' which, we argue, is 
qualitatively different from the so-called rejuvenation effect 
observed just after the $T$-shift.
}

\kword{spin glass, slow dynamics, aging, droplet theory}

\begin{document}
\sloppy
\maketitle

\section{Introduction}

Recently, in studies on slow dynamics in spin
glasses~\cite{rev1,rev2,rev3}, the apparently contradictory 
phenomena, i.e., {\it rejuvenation} (or {\it chaos}) and {\it memory} 
effects in aging dynamics, have been intensively 
investigated~\cite{JVHBN,revBouch}. 
In fact the phenomena were already observed in the early stage of 
study on aging in spin glasses by the so-called temperature-cycling 
protocol.~\cite{rejuvenat1}. In the protocol we quench a spin glass to 
a temperature, say $T_1$, below the spin-glass (SG) transition 
temperature $\Tc$ from above it and let the system equilibrate (or age) 
for a period of $\twone$. Subsequently, we change the temperature to 
$T_2 \ (<T_1)$, age the system for a period of $\twtwo$, and then the 
temperature is turned back to $T_1$. For a certain range of the 
parameters $T_1, T_2, \twone$ and $\twtwo$ some quantities such as the
out-of-phase component of ac susceptibility, $\chitwo$, exhibit the 
following behavior. Just after the first $T$-shift $\chitwo$ behaves as 
if the system were quenched to $T_2$ directly from above $\Tc$, or it 
looks having forgotten the aging at $T_1$ before the $T$-shift. This is 
called the rejuvenation (or chaos) effect. However, after the 
temperature is turned back to $T_1$, $\chitwo$ observed is the one we 
expect as a simple extension of $\chitwo$ already aged by $\twone$ at 
$T_1$. Thus the system definitely preserves the memory of the previous 
aging at $T_1$, while it apparently exhibits the rejuvenation behavior, 
in the aging process at $T_2$. The proper understanding of such a 
peculiar phenomenon is believed to shed light not only on the mechanism 
behind the aging dynamics but also the nature of the SG phase itself. 
Furthermore it will provide us with powerful concepts to 
understand the glassy dynamics in various related systems such as 
orientational glasses,~\cite{o-glass} polymers,~\cite{polymer} and 
interacting nanoparticles systems.~\cite{nano-p}

By the real-space interpretation, or by the droplet 
picture~\cite{BM-chaos,FH-88-EQ,FH-88-NE}, which we have been adopting 
in our recent studies~\cite{ours1,ours2,ours4,Lorenzo-00}, the SG order is 
considered to grow up slowly in aging processes. In particular, we have 
demonstrated that the SG coherence length, which we regards as the mean
size of SG domains developed in aging, continuously grows even under the 
$T$-shift process.~\cite{ours4} This we call the {\it cumulative memory} 
effect. We have further extended this characteristics to a scenario that 
the SG short-range order ever grows continuously with growth rates 
sensitively dependent on the temperature so long as the system is in the 
SG phase.~\cite{Lorenzo-00} Let us call this the {\it cumulative memory 
scenario}, and denote the mean size of SG domains as $R_{T[t]}(t)$, where 
$T[t]$ symbolically represents the temperature changes that the system 
has experienced up to time $t$ from the first quench to the SG phase at 
$t=0$. It has been demonstrated that the time evolution of 
zero-field-cooled (ZFC) magnetizations observed in various schedules of
temperature changes but with a common initial quench 
condition~\cite{Ito-00} are well described by a unique function of 
$R_{T[t]}(t)$.~\cite{Lorenzo-00}  

The purpose of the present work is to numerically explore to what extent 
the simulated data of $\chitwo$ in the $T$-shift protocol of the 3D Ising 
EA SG model are compatible or incompatible with the cumulative memory 
scenario, and with the rejuvenation (chaos) and memory effects mentioned 
above. For this purpose we certainly need a few more length and time 
scales than $R_{T[t]}(t)$. One is $R_{T[t]}(t)$ at $t=\twone$, i.e., the 
mean domain size grown in the isothermal aging after quench to the SG
phase, which is denoted as $R_{T_1}(\twone)$. After the temperature is
shifted to $T_2$ spin configurations within each domains, which were in
local equilibrium of $T_1$ just before the $T$-shift, gradually become
in local equilibrium of $T_2$. By the word `gradually' we mean
that the change associates with slowly growing droplets (or subdomains) 
of a mean size $L_{T_2}(\tau) \ (< R_{T_1}(\twone))$ with $\tau = 
t-\twone$. Here we call this the {\it droplets-in-domain} scenario 
(previously called the {\it quasi-domains-within-domains}
picture~\cite{ours4}) . 

A key quantity of the present work is the time scale required for 
$L_{T_2}(\tau)$ to catch up $R_{T_1}(\twone)$.~\cite{ours4} At time
scales larger than this one, denoted as $\twoneeff$, behavior of physical
quantities such as $\chitwo$ cannot be distinguished, within the
accuracy of measurement, from the corresponding behavior in the
isothermal aging at $T_2$. In other words, the system merges to a 
$T_2$-isothermal aging state at the time scale of $\twoneeff$ after the
$T$-shift. The latter is regarded as the {\it effective waiting time} of
the $\twone$-aging at $T_1$ reread as an $T_2$-isothermal aging.
If the cumulative memory scenario holds, $\twoneeff$ is the time
required for the SG coherence to grow in the $T_2$-isothermal aging up
to $R_{T_1}(\twone)$, i.e., 
\begin{equation}
        L_{T_2}(\twoneeff) = R_{T_2}(\twoneeff) = R_{T_1}(\twone).
\label{eq:tweff}
\end{equation}
is expected to hold. Here the first equality simply indicates that 
$L_T(t)$ has the same functional form as that of $R_T(t)$ since the both 
growth processes are governed by common thermally-activated dynamics. 

In the present work we have extensively examined $T$-shift processes in 
the 3D Gaussian EA model with $\Tc \simeq 0.95J$~\cite{Tc3DJ} through 
the ac susceptibility $\chitwo$ simulated by the standard Monte Carlo 
(MC) simulation. Here $J$ is the variance of the interactions. The 
temperature range investigated is $T \sim [0.4, 0.7]$ in unit of $J$, 
and the time range is up to $10^5$ MC steps. One of the results we have 
found is that in negative (positive) $T$-shift protocols with $T_1 
\rightarrow (\leftarrow)\  T_2$  Eq.(\ref{eq:tweff}) (the same equation 
but with the suffix 1, 2 interchanged) holds well when $\Delta T = 
T_1 - T_2 = 0.1$.  This confirms the cumulative nature of aging in both 
negative and positive $T$-shift protocols with a small $\Delta T$. A more 
interesting result is that, for the negative $T$-process with $T_1=0.7, 
T_2=0.4$ significant violation of Eq.(\ref{eq:tweff}) has been observed, 
while in the corresponding positive $T$-shift protocol 
Eq.(\ref{eq:tweff}) is satisfied within accuracy of the present 
numerical analysis. The former non-cumulative memory effect has been, 
for the first time to our knowledge, observed in simulations on the 
3D EA models. Although the phenomena appear asymmetrically with respect 
to the direction of temperature changes, the deviation from the 
cumulative memory scenario in the negative $T$-shift appears 
qualitatively similarly to the one observed recently in 
experiments on the AgMn spin glass.~\cite{JYN} We tentatively regard them 
as an effect associated with the temperature-chaos effect in the 
equilibrium SG phase,~\cite{BM-chaos,FH-88-EQ} and call them the 
{\it `chaos effect'}. In contrary, as in the previous work,~\cite{PRR,BB} 
such rejuvenation effects observed experimentally just after the $T$-shift 
have not been detected even with $\Delta T =0.3$ in the present work.  

The organization of the paper is as follows. In the next section we 
describe our strategy of the simulation and the method to evaluate 
$\chitwo$ from the spin autocorrelation function making use of the 
fluctuation-dissipation theorem. In \S 3 we explain how to specify the 
effective waiting time $\twoneeff$ from the obtained data of $\chitwo$, 
and present the results of $\twoneeff$, or $R_{T_2}(\twoneeff)$ 
for $T$-shift processes with various sets of $T_1$ and $T_2$. In the
final section we discuss our results, emphasizing on the length and time
scales involved as well as on the relation to the experimental results.

\section{Method of Analysis}

Numerical simulation on a well-defined microscopic SG model, such as the
EA model investigated in the present work, is of quite importance in
studying aging phenomena. It enables us to faithfully realize any
$T$-shift process and observe any quantity in principle.  For example,
the SG coherence length $\xi_{T[t]}(t)$, which we regard as the mean SG
domain size $R_{T[t]}(t)$, has been calculated from the replica overlap 
function.~\cite{ours4} For the isothermal process of the 3D Ising 
Gaussian EA model which we study in the present work, it is well
described by the power law~\cite{ours4,Kisker-96,Marinari-98-VFDT}
written as  
\begin{equation}
        R_T(t)/l_0 = b_T(t/t_0)^{1/z(T)},
\label{eq:RT(t)}
\end{equation}
where $l_0$ and $t_0$ are microscopic length and time scales ($l_0=$ 1
lattice distance and $t_0=$ 1 MC step for simulated results), $b_T$ is a 
weakly $T$-dependent constant, and the exponent $1/z(T)$ linearly
depends on $T$ except for the region near $\Tc$~\cite{ours1}. 

The growth law of $R_{T}(t)$ in isothermal aging different from 
Eq.(\ref{eq:RT(t)}) was proposed in the droplet theory due to Fisher and 
Huse.~\cite{FH-88-NE} It is written as~\cite{YHT} 
\begin{equation}
 R_T(t)/L_0(T)=\tilde{L}(t/\tau_0(T)),
\label{eq:LTcross}
\end{equation}
with the scaling function $\tilde{L}(x)$ given by 
\begin{eqnarray}
\tilde{L}(x) \sim \left\{ 
\begin{array}{cc}
 x^{1/z}  & (x \ll 1), \\
 \log^{1/\psi}(x) & (x\gg 1). 
\end{array}
\right.
\label{eq:scal-g}
\end{eqnarray}
Here $L_0(T)\ (\sim l_0\epsilon^{-\nu})$ is the crossover length,
$\tau_0(T) \ (\sim t_0\epsilon^{-z\nu})$ is the attempt time for thermal
activation process of droplets, where $\epsilon=(\Tc-T)/\Tc$, and $z$
and $\nu$ are the critical exponents associated with the criticality at
$\Tc$. The exponent $\psi$ in Eq.(\ref{eq:scal-g}) is, on the other hand,
intrinsic to the activation dynamics of droplets. We have recently
confirmed the above growth law, including the crossover from the
critical dynamics ($x\ll 1$) to the activated dynamics  ($x\ll 1$), in
the EA SG model but in 4 dimension~\cite{YHT} (see also
\cite{BB}). However, the simulated data of $R_T(t)$ in the 3D EA model
are compatible with both the power law of Eq.(\ref{eq:RT(t)}) and the
logarithmic law in Eq.(\ref{eq:scal-g}).~\cite{BB,Kisker-96} 

Our strategy in the present work is as follows. Because of the
circumstances of the 3D EA model mentioned just above as well as those
of the recent experiment,~\cite{JYNKI} we do not go into
the question which growth law is a correct one for $R_T(t)$ observed in
the time-window ($\nle 10^5$ MCS) of the simulations, and simply use our
results, i.e., Eq.(\ref{eq:RT(t)}), in relating a time scale of
observation to a length scale of the SG short-range order. Then we
examine whether the cumulative memory scenario is sufficient or not to 
interpret the obtained results of the length scales. We also restrict 
ourselves to the temperature range of $T/J=[0.4,0.7]$ as already noted 
in \S 1. In this temperature range $1/z(T)$ in Eq.(\ref{eq:RT(t)}) is
well proportional to $T$ and the dynamics is considered to be dominated
by the activated process. But the prefactor $b_T$ still exhibits weak
dependence on $T$ even in this range which will turn out not to be 
neglected in our present analysis. Lastly we have examined $T$-shift
processes with various values of the temperature differences $\Delta T$,
in particular, a larger one than that studied in our previous 
work.~\cite{ours4}

In the aging study at $T$ through the ac susceptibility at frequency 
$\omega$ we need to introduce another length scale, $L_T(\tomega)$, 
with $\tomega=2\pi/\omega$. It is a mean size of spin clusters (or 
droplets) which can respond to the ac field at $T$. In the droplet 
picture the aging (or $t$-dependent) part of $\chitwo$ in an isothermal
aging is described by a function of
$L_T(\tomega)/R_T(t)$.~\cite{FH-88-NE,ours2} As will be discussed in \S4
below, $\chitwo$ in some $T$-shift process to $T$ at $t=\tw$ is given in
terms of $L_T(\tomega)/L_T(\tau)$ where $\tau = t -\tw$. Thus it
provides us information of $R_T(t)$ or $L_T(\tau)$ in the
aging process since $L_T(\tomega)$ is independent of $t$ or $\tau$.  

By the experimental condition of measuring $\chitwo$, $t$ or $\tau$ is
necessarily larger than $\tomega$ in general. This time regime is called
the quasi-equilibrium one, where the fluctuation-dissipation theorem
(FDT) is expected to hold well, though
approximately.~\cite{AlbaHOR-87,RefregierOHV-88}  Therefore, in the 
present work, the out-of phase component of ac susceptibility, $\chitwo$, 
is evaluated from the spin auto-correlation function
\begin{equation}
        C(\tau; t) = \overline{ \langle S_i(\tau+t)S_i(t)\rangle }, 
\label{eq:def-Cor}
\end{equation}
via the FDT as~\cite{ours2} 
\begin{equation}
     \chitwo \simeq  - \left. {\pi \over 2T}
    {\partial \over \partial {\rm ln} \tau}
     C(\tau; t)\right|_{\tau = \tomega}.
\label{eq:chi-C-1}
\end{equation}
In Eq.(\ref{eq:def-Cor}), $S_{i}(t)$ is the sign of the Ising spin at site 
$i$ at time $t$ which is measured in unit of one MC step. The 
over-line denotes the averages over sites and over different realizations 
of interactions (samples), and the bracket the average over thermal noises 
(or different MC runs). In this evaluation of $\chitwo$ we are completely  
free from any nonlinear effect of the ac-field amplitude.~\cite{PRR}

In our previous work~\cite{ours4} we studied the susceptibility defined by
\begin{equation}
   \chitilde = \left. {1 \over T}[1-C(\tau; t)]\right|_{\tau = \tomega}.
\label{eq:chi-tilde}
\end{equation}
It is just the ZFC susceptibility: the induced magnetization 
(divided by $h$) at an elapsed time of $\tau$ under the field $h$ which is 
switched on after the system has aged under $h=0$ by a period of $t$.  For 
slow processes of our present interest, $\chitilde$ is essentially 
regarded as the in-phase component of the ac susceptibility, $\chione$.
As in the experiments, simulated $\chitwo$ exhibits larger effects of 
aging relatively to its own absolute magnitude than $\chione$ or 
$\chitilde$ does. However we have to numerically evaluate the logarithmic 
derivative in Eq.(\ref{eq:chi-C-1}) to estimate $\chitwo$. In the present 
work we have calculated several $C(\tau; t)$ in Eq.(\ref{eq:def-Cor}), 
each of which is the average over one MC run for each sample but
typically over 1600 samples. The linear system size is fixed to
$L=24$. The error bars on $\chitwo$ drawn in the figures shown below
indicate the variance in the results of the numerical derivative on
these several sets of $C(\tau; t)$.

\section{Results}

\subsection{Isothermal aging}

\begin{figure}
\leavevmode\epsfxsize=80mm
\epsfbox{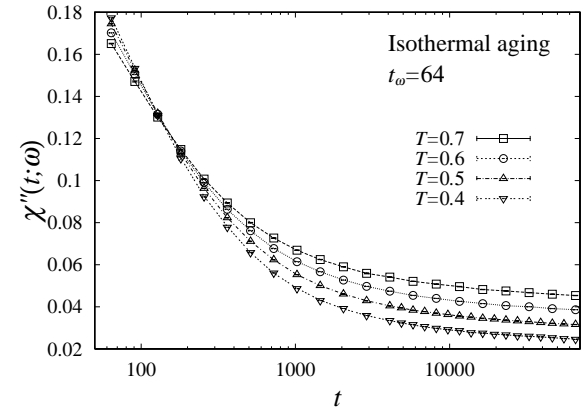}
\caption{$\chitwo$ with $\tomega=64$ in the isothermal aging at various 
temperatures.}
\label{fig:isoT_64}
\end{figure}

Before going into discussions on the $T$-shift protocol, let us here 
present the results of $\chitwo$ obtained in the isothermal aging. In 
Fig.~\ref{fig:isoT_64} we show $\chitwo$ with $\tomega=64$ in the 
isothermal aging at various temperatures. They play an important role 
in the following arguments on the $T$-shift protocol, and we call 
them the {\it reference curve} at each temperature. 

\begin{figure}
\leavevmode\epsfxsize=80mm
\epsfbox{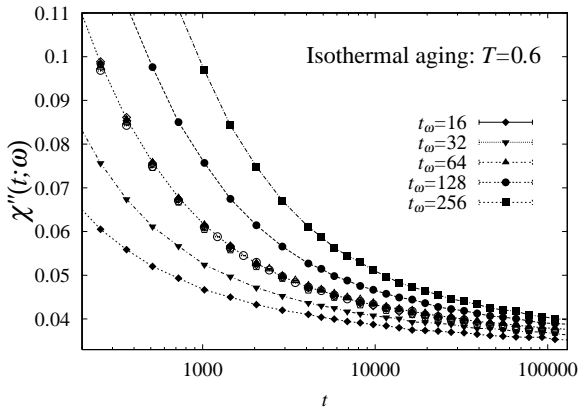}
\caption{$\chitwo$ in the isothermal aging at $T=0.6$. The filled symbols 
are the data plotted directly against $t$, while the open symbols are
 the same data plotted against  $64 t/t_\omega$.}
\label{fig:iso-T6}
\end{figure}

In Fig~\ref{fig:iso-T6} $\chitwo$ at $T=0.6$ for $\tomega=16 \sim 256$ 
are shown. The filled symbols are raw data plotted directly against $t$, 
while the open symbols are the same $\chitwo$ plotted against 
$\omega t$ with no vertical shifts of the data sets. All the sets of
data thus plotted nicely collapse to a universal 
curve. This reconfirms that the $\omega t$-scaling of $\chitwo$ holds 
also in the present model spin glass~\cite{PRR} as observed 
experimentally~\cite{rev1}. As pointed out in \S 2, the time 
evolution of $\chitwo$ is considered to be a function of
$\Lomega/R_T(t)$ in the droplet picture. 
The $\omega t$-scaling then comes out from the first equality 
of Eqs.(\ref{eq:tweff}) and (\ref{eq:RT(t)}).~\cite{ours2}. The response 
in equilibrium, $\chitwoeq = \lim_{t \rightarrow \infty}\chitwo$, is 
hardly extracted from our present data. We could not detect even its
relative difference with $\omega$, which should be reflected as the
vertical shifts of the data sets in the above scaling analysis. 

\subsection{$T$-shift protocol}

\begin{figure}
\leavevmode\epsfxsize=80mm
\epsfbox{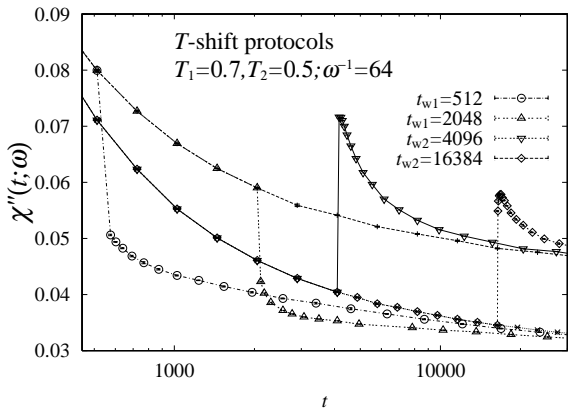}
\caption{$\chitwo$ in aging with negative and positive $T$-shifts
 between $T_1=0.7$ and $T_2=0.5$ at $t=t_{{\rm w}i}$ indicated in the
 figure. The upper and lower curves with the smaller 
 symbols represent the $T_1$- and $T_2$-reference (isothermal) curves,
 respectively. 
}
\label{fig:Tshf-dnup}
\end{figure}

In Fig.~\ref{fig:Tshf-dnup} we show $\chitwo$ with $\tomega=64$
numerically observed in the {\it negative (positive)} $T$-shift protocol. 
The temperature is changed from $T_1=0.7 \ (T_2=0.5)$ to 
$T_2=0.5 \ (T_1=0.7)$ at different waiting times $\twone\ (\twtwo)$. The
observation starts from $t=t_{{\rm w}i}+t_\omega$ after each $T$-shift. 
Similarly to $\chitilde$ previously investigated~\cite{ours4}, we see 
the following characteristic features. 
\begin{enumerate}
\renewcommand{\labelenumi}{\alph{enumi})}
\item Each $\chitwo$ rapidly undershoots (overshoots) the 
$T_2\ (T_1)$-reference curve.  
\item $\chitwo$ merges to the $T_2\ (T_1) $-reference curve from 
below (above). 
\end{enumerate}
For the negative $T$-shift, in particular, the value of $\chitwo$ just
after the negative $T$-shift is the larger relatively to the
$T_2$-reference curve, the larger is $\twone$. However, $\chitwo$ just
after the shift does not exhibit overshooting of the $T_1$-reference
curve, a phenomenon which we call the {\it strong rejuvenation} effect
in the present paper. This is also the case even if we dare to measure
$\chitwo$ at $\tau=t-\twone$ smaller than $\tomega$.~\cite{BB}
 
\subsection{Effective waiting time}

\begin{figure}
\leavevmode\epsfxsize=80mm
\epsfbox{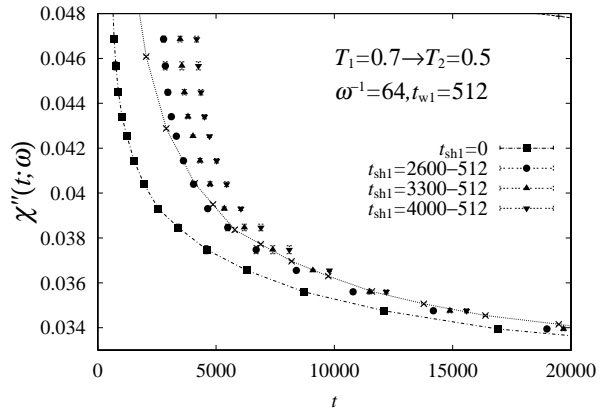}
\caption{$\chitwo$ in the negative $T$-shift protocol from $T_1=0.7$ to 
$T_2=0.5$. Each symbol represents the branch of $\chitwo$ at $t>\twone$
 shifted by an amount of $\tshone$ indicated in the figure. the line 
with the smaller symbols is the $T_2$-reference curve.
}
\label{fig:Tshf75}
\end{figure}

Feature b) above is examined in more details in Figs.~\ref{fig:Tshf75}
and \ref{fig:Tshf57}. We note that the $t$-axis in these figures
is linear in $t$.  Within the time window of Fig.~\ref{fig:Tshf75} the
merging of bare $\chitwo$ (denoted by $\tshone=0$) to the $T_2$-reference 
curve is not seen. If, however, the branch of $\chitwo$ at $t>\twone$ is 
shifted to the {\it right} by an amount denoted by $\tshone$, it crosses 
the reference curve and merges to it at a smaller $t$ than that with
$\tshone=0$.  At a certain value 
of $\tshone$ ($\simeq 3300-512$ in the figure) it merges to the reference
curve and lies on it afterwards. We regard time $\tau$ required for the 
shifted branch to merge to the reference curve in this situation as the 
effective waiting time, $\twoneeff$, introduced in \S 1. If $\tshone$ is 
further increased the branch merges to the reference curve from above and 
again at larger $t$ than $\twoneeff$. Thus the chosen $\tshone$ which
corresponds to $\twoneeff$ yields the shortest time for the shifted
branch to merge to the reference curve.  

An interesting observation in the above analysis is that the time 
at which the properly shifted branch merges to the 
$T_2$-reference curve is nearly equal to $2\twoneeff$; 
$\twone+\tshone+\twoneeff \simeq 2\twoneeff$ and so 
$\twoneeff \simeq \twone + \tshone$. This aspect, which is by no means 
trivial, has been commonly observed in most of the $T$-shifts examined 
in the present work. Using this observation, we estimate errors of
$\twoneeff$ as  
follows. We judge by eyes the largest $\tshone$ for which the shifted 
branch of $\chitwo$ certainly crosses with but not merges to the 
reference curve, and this value of $\tshone$ gives a smallest estimate 
of $\twoneeff$. Similarly the smallest $\tshone$ for which the shifted 
branch merges to the reference curve at $t \nge 3\twoneeff$ yields a 
largest estimate of $\twoneeff$. The three shifted branches shown in 
Fig.~\ref{fig:Tshf75} correspond to these smallest, mean, and largest 
estimates for $\twoneeff$. 

The above analysis for the negative $T$-shift protocol also works for
the positive  $T$-shift protocol. A typical example from $T_2=0.5$
to $T_1=0.7$ with $\twtwo=16384$ is shown in Fig.~{\ref{fig:Tshf57}, for 
which we have to reread the suffix 1 by 2 and vice versa in the above 
argument. Also in this case the branch of
$\chitwo$ at $t>\twtwo$ is shifted to the {\it left} by $\tshtwo$. A too
large $\tshtwo$ makes the shifted branch to overshoot the $T_1$-reference
curve, while a too small $\tshtwo$ significantly delays the merging. With 
a proper chosen $\tshtwo$ we obtain $\twtwoeff\ (\simeq 1600$ from the 
figure) which satisfies  $\twtwoeff = \twtwo - \tshtwo$. Its error bar is 
similarly evaluated from the two other $\tshtwo$'s indicated in the figure. 

\begin{figure}
\leavevmode\epsfxsize=80mm 
\epsfbox{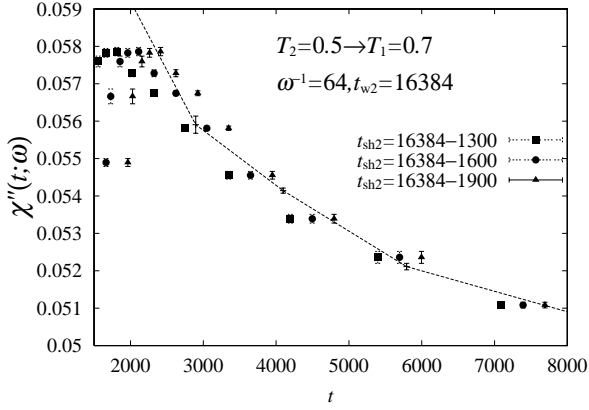}
\caption{$\chitwo$ with $\tomega=64$ in the positive $T$-shift protocol from
 $T_2=0.5$ to $T_1=0.7$ with $\twtwo=16384$. The data in a large time scale 
 are shown in Fig.~\ref{fig:Tshf-dnup}. The three sets of symbols
 represent the branches of $\chitwo$ at $t>\twtwo$ shifted to the left
 by the amount $\tshtwo$ indicated in the figure. The line is the 
$T_1$-reference curve.
}
\label{fig:Tshf57}
\end{figure}

\subsection{Cumulative memory and `chaos' effects}

In Fig.~\ref{fig:Efftw} we plot $\twoneeff$ ($\twtwo$) as a function
of $\twone$ ($\twtwoeff$) obtained in the previous subsection in the
negative (positive) $T$-shift protocol for three sets of ($T_1,
T_2$). Here we have followed the idea of `twin-experiments' in the
recent work.~\cite{JYN} Before the explanation of the lines drawn in the 
figure, we note that the data points of both negative and positive 
$T$-shifts with $T_1=0.7, \ T_2=0.6$ are seen to lie on a certain common 
curve, while this is not the case for those with $T_1=0.7$ and $T_2=0.4$. 
The former is expected from the cumulative memory scenario. But the latter 
data points clearly indicate a violation to the scenario irrespectively of 
the growth law of the SG domains. 

Now let us explain the lines in Fig.~\ref{fig:Efftw}. The solid ones
represent the relation between the  $\twoneeff$ ($\twtwo$) and $\twone$
($\twtwoeff$) when the cumulative memory scenario represented by 
Eq.(\ref{eq:tweff}) (the one whose suffix 1, 2 interchanged) combined
with the growth law of Eq.(\ref{eq:RT(t)}) holds. For the latter we have
explicitly used the following sets of the parameter values 
$(T; z(T), b_T)$ we previously obtained:~\cite{ours1} (0.7; 8.71, 0.779), 
(0.6; 9.84, 0.782), (0.5; 11.76, 0.800) and (0.4; 14.80, 0.818). 
As seen in the figure, for a small $\Delta T \ (=0.1)$, both $\twoneeff$
and $\twtwoeff$ lie on the solid line. The results confirm the
cumulative memory scenario described in \S 1. With $\Delta T =0.2$,
$\twtwoeff$ still lie on the solid line but $\twoneeff$ tends to
deviate, though a little, from it. For $\Delta T = 0.3$, $\twoneeff$ 
significantly deviate from the solid line, which is incompatible with 
the cumulative memory scenario. It should be emphasized, however, that
the corresponding $\chitwo$ (not shown) in this process exhibit 
features a), b) mentioned in \S 3.2, i.e., no strong rejuvenation.

If the weak $T$-dependence of $b_T$ is discarded the condition of 
Eq.(\ref{eq:tweff}) is reduced to  
\begin{equation}
   \left({\twoneeff \over \tau_0}\right) = 
\left({\twone \over \tau_0}\right)^{T_1/T_2},
\label{eq:cond-tweff-appro}
\end{equation} 
which is also shown by the dotted lines in Fig.~\ref{fig:Efftw}.
For $\Delta T = 0.1$ the effect of the $T$-dependence of $b_T$ is
negligibly small. The effect is, however, significant for $\Delta T \ge
0.2$. Thus the $T$-dependence of $b_T$ has to be properly taken into
account to judge the cumulative nature of memory observed even in the
temperature range examined in the present simulation. 

\begin{figure}
\leavevmode\epsfxsize=80mm
\epsfbox{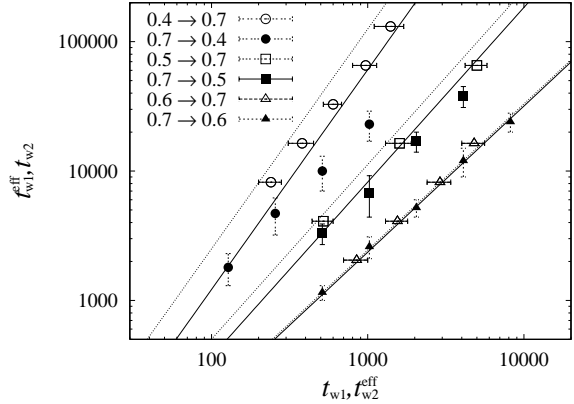}
\caption{$t_{{\rm w}i}^{\rm eff}$ vs $t_{{\rm w}i}$ for the
 negative ($i=1$) and positive ($i=2$) $T$-shift protocols for
 three sets of $T_1$ and $T_2$ indicated in the figure. The lines
 represent the expected behavior from Eqs.(\ref{eq:tweff}) and
 (\ref{eq:RT(t)}) as explained in the text.
}
\label{fig:Efftw}
\end{figure}
\begin{figure}
\leavevmode\epsfxsize=80mm
\epsfbox{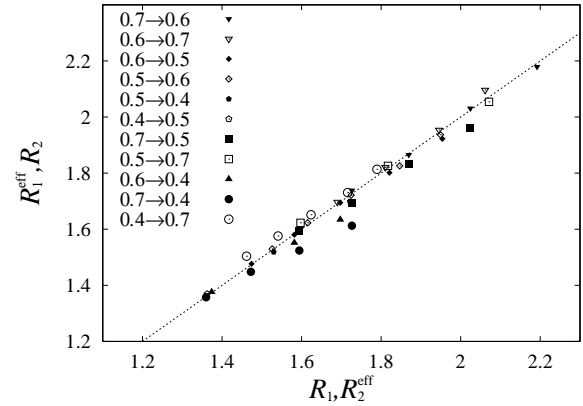}
\caption{Relations $t_{{\rm w}i}^{\rm eff}$ vs. $t_{{\rm w}i}$ drawn by 
means of the corresponding domain sizes evaluated by Eq.(\ref{eq:tweff})
with Eq.(\ref{eq:RT(t)}). Here we omit the error bars. For all the sets
 of $T_1, T_2$ their magnitudes are comparable with those shown in Fig. 6.  
}
\label{fig:EffR}
\end{figure}

In Fig.~\ref{fig:EffR} we replot our data in Fig.~\ref{fig:Efftw} as 
well as  those of other sets of ($T_1, T_2$) in terms of the lengths, 
where $R_1 \equiv R_{T_1}(\twone)$ and $R_1^{\rm eff} \equiv 
R_{T_2}(\twoneeff)$ which are evaluated by Eq.(\ref{eq:RT(t)}) using 
$\twone$ and $\twoneeff$ extracted in the negative $T$-shift 
processes ($R_2$ and $R_2^{\rm eff}$ in the positive $T$-shift process 
are similarly evaluated). The line in the figure is what is expected from 
the cumulative memory scenario, i.e., $R_i^{\rm eff}=R_i$ for both $i=1,2$. 
We see clearly that this is the case for both negative and positive 
$T$-shift processes with $\Delta T=0.1$ within the time window of 
the present simulation. For $\Delta T=0.2$  the data of the positive 
$T$-shift satisfy the condition $R_2^{\rm eff}=R_2$, but those of the 
negative $T$-shift exhibit the tendency $R_1^{\rm eff} < R_1$. Behavior of 
the $T$-shift with $\Delta T= 0.3$, i.e., $T_1=0.7$ and $T_2=0.4$ is as 
already described above and is interpreted below to be due to the `chaos 
effect.' 

According to the theory for the temperature-chaos in spin 
glasses~\cite{FH-88-EQ,BM-chaos}, the SG equilibrium configurations at 
different temperatures, $T_1$ and $T_2$, are completely uncorrelated with 
each other in the length scale larger than $l_{\Delta T} \propto \Delta 
T^{-1/\zeta}$, where $l_{\Delta T}$ is called the overlap length and 
$\zeta \ (> 0)$ the chaos exponent. An important problem here is how the
existence of $\lovlp$ affects the non-equilibrium aging 
dynamics. Let us consider a negative $T$-shift process with $\Delta T$, 
for which $\lovlp$ is supposed to be sufficiently smaller than 
$R_{T_1}(\twone)$, and introduce the time scale $t_{\rm ov1}$ by 
$L_{T_2}(t_{\rm ov1})=\lovlp$. At a time range after the $T$-shift 
specified as $\twoneeff \gg \tau \gg t_{\rm ov1}$, a longer part of the
memory imprinted before the $T$-shift is still preserved, but such a
memory is expected to be irrelevant to the equilibration 
process to the SG ordered state at $T=T_2$. Thus the system 
looks as if it is already in the isothermal aging state at $T_2$. 
Then, if our analysis to determine $\twoneeff$ is applied to 
this $T$-shift process, the expected result is 
$\twoneeff \simeq t_{\rm ov1}$ irrespectively of $\twone$.  
The circumstances are the same for the positive $T$-shift protocol. 
Consequently, $R_i^{\rm eff}$ in Fig.~\ref{fig:EffR} is expected to
saturate to $\lovlp$ at large $R_i$ both for $i=1,2$, and the data for
the negative and positive $T$-shifts come out symmetrically with respect
to the line of $R_i^{\rm eff}=R_i$. 

We tentatively interpret our data of the negative $T$-shift with 
$\Delta T=0.3$ as an early stage of the saturation described above. 
Unfortunately, the data are so limited that we cannot figure out a 
value of $\lovlp$. Also the corresponding positive $T$-shift data 
nearly coincide with the line $R_i^{\rm eff}=R_i$, i.e., the two sets 
of data are by no means symmetric with respect to the line. One of 
the reason of this asymmetric behavior may be due to our method to
specify the effective aging time combined with the time scales 
in our simulation. Although $\lovlp$ is common to the negative and 
positive $T$-shifts, the separation of time scales $\twtwo, 
t_{\rm ov2}$ and $\twtwoeff$ in the positive $T$-shift is much smaller 
than that of $\twone, t_{\rm ov1}$ and $\twoneeff$ in the negative 
$T$-shift. This is due to a large difference in the growth rates at 
the two temperatures. It is then rather hard to detect a possible small 
difference between $t_{\rm ov2}$ and $\twtwoeff$ within our present
analysis. With these reservations, we interpret our results of the
$T$-shift process with $\Delta T = 0.3$ as a dynamic process which
reflects the temperature-chaos predicted for the equilibrium SG phase.

\section{Discussions}

In the $T$-shift protocol examined in the present study, the cumulative 
memory scenario has been confirmed for $T$-shift processes with a small 
magnitude of the shift, i.e., $\Delta T = T_1 - T_2 = 0.1$. This has
been done by close comparisons of $\chitwo$ after the $T$-shift with
that in the isothermal aging at $T_2$ (reference curve). However, there
have been little experiments which directly measure $\twoneeff$
similarly to our analysis.~\cite{Mamiya,VD} An example is the one by
Mamiya {\it et al},~\cite{Mamiya} who analyzed the aging dynamics in the
SG-like phase of a ferromagnetic fine particles system. In the $T$-shift
process with $T_1=49$K and $T_2=47$K (with $T_{\rm g}\simeq 70$K) they
observed $\twoneeff \sim 3 \times \twone$ for $\twone=2.0,...,15.0$ks. 
If we suppose $\tau_0=10^{-6}$s for a magnetic moment carried on by each
fine particle,~\cite{Mamiya-pr} we obtain $\twoneeff \simeq 
(2.4 \sim 2.7)\times \twone$ from Eq.(\ref{eq:cond-tweff-appro}). The
result is rather satisfactory and implies that the cumulative memory
scenario works as well for the $T$-shift process with a small $\Delta T$
in this SG material. 

Our results on the negative $T$-shift protocol with $\Delta T = 0.3$ have 
turned out to be incompatible with the cumulative memory scenario. The 
period $\twoneeff$ necessary for the system to become in a 
$T_2$-isothermal aging state after the $T$-shift is significantly smaller 
than the value of $\twoneeff$ estimated from Eq.(\ref{eq:tweff}) combined 
with Eq.(\ref{eq:RT(t)}). In this negative $T$-shift process which
violates the cumulative memory scenario, however, the strong rejuvenation 
phenomenon just after the $T$-shift, which is described in \S 3.2, has 
not been detected. We have therefore attributed the non-cumulative memory 
effect we have found to the `chaos effect'.

Quite recently J\"onsson {\it et al.} (JYN) have reported the chaos effect 
which symmetrically appears in the positive and negative $T$-shifts in a 
Heisenberg-like spin glass AgMn.~\cite{JYN} They have measured the ZFC 
magnetization with schedules of temperature change corresponding to the 
$T$-shift protocol discussed in the present work but within a very small
range of $\Delta T\ (\le 0.012\Tc)$. The logarithmic-time 
derivative of the ZFC magnetization, $S(t;\twone)$, exhibits a peak, whose 
position is considered to be at $\tau=t-\twone \simeq \twoneeff$, 
the time required for the merging to an isothermal state at the new 
temperature just investigated in the present work (see the discussion 
below). In fact, our $R_1^{\rm eff}$-vs-$R_1$ plot in Fig.~\ref{fig:EffR} 
of the negative $T$-shift with $\Delta T = 0.3$ is in qualitative agreement
with their $L_{\rm eff}$-vs-$L_{T_i}(\tw)$ plot, where their $L_{\rm eff}
\ (L_{T_1}(\tw))$ just corresponds to our $R_1^{\rm eff}\ (R_1)$. In 
contrast to our numerical observation, however, their data for the positive 
$T$-shift plotted in our way appear symmetrically to the negative one 
with respect to the line $R_i^{\rm eff}=R_i$. The overlap length $\lovlp$ 
estimated by scaling analysis has turned out to be larger than 
$R_i\ (=L_{T_i}(\tw))$, or before the saturation of $R_i^{\rm eff}$
mentioned in \S 3.4.  One of the reasons of this discrepancy between
their experiment and our simulation may be the Heisenberg-spin nature in
their material AgMn; Heisenberg spin glasses are more chaotic than Ising 
spin glasses.~\cite{JYN,DVBHIK} This point is very interesting and to be 
further pursued. 

Next let us make a few comments on aging at a time range $\tau = t-\twone 
\nle \twoneeff$ after the $T$-shift, which we have called 
the transient regime of the $T$-shift process.~\cite{ours4} A main idea 
for this regime is the droplets-in-domain scenario which involves at 
least two characteristic length scales as mentioned in \S~1. One is the 
mean domain size at $\tau=0$, i.e., $ R_{T_1}(\twone)$ and the other is 
$L_{T_2}(\tau)$, the mean size of droplets (or subdomains)  which are 
already in local equilibrium of the shifted temperature $T_2$ at time 
$\tau$ after the $T$-shift. Associated with the growth of $L_{T_2}(\tau)$ 
some peculiar features have been observed. An example is a non-monotonic 
time evolution of the energy density in a positive $T$-shift process (see 
Fig. 4 in~\cite{ours4}). It is recently named as the Kovacs effect 
in~\cite{BB} since the qualitatively similar phenomenon was first 
observed in polymer glasses.~\cite{Kovacs} 

In the droplets-in-domain scenario here we implicitly assume that
droplets (or subdomains) in local equilibrium of $T_2$ distinguish
themselves from those in local equilibrium of $T_1$. On the other hand,
the Kovacs effect has been observed in negative $T$-shift processes with
$\Delta T = 0.2$, for which the `chaos effect' is not clearly detected in
Fig.~\ref{fig:EffR}, or $\lovlp > R_{T_1}(\twone)$. This strongly
suggests that in nonequilibrium aging dynamics the spin configurations
which we have so far supposed to be in local equilibrium at two
different temperatures differ from each other even in length scales 
shorter than the equilibrium $\lovlp$ of the corresponding temperatures. 
This viewpoint is in contrast to the argument of the temperature-chaos
in equilibrium, and is worthy to be investigated.

From the droplets-in-domain scenario mentioned just above, the strong
rejuvenation experimentally observed in the ac susceptibility
measurement discussed in \S 3.2 can be regarded as one 
of such peculiar phenomena in the transient regime. As mentioned in \S 2, 
the ac response associates another short length scale $L_{T_2}(\tomega)$. 
In the time range $\tomega \ll \tau \ll \twoneeff$ so that 
$L_{T_2}(\tomega) < L_{T_2}(\tau) < R_{T_1}(\twone)$ holds, the 
$t$-dependent part of $\chitwo$ is governed dominantly by the ratio 
$L_{T_2}(\tomega)/L_{T_2}(\tau)$ since droplets responding to the ac
field less feel the existence of larger domains of $R_{T_1}(\twone)$
than that of smaller subdomains of $L_{T_2}(\tau)$. The strong
rejuvenation alone is therefore not necessarily incompatible with the
cumulative memory scenario. In neither the previous~\cite{ours4,PRR,BB}
nor the present simulations, however, the strong rejuvenation has been
detected. This may be again attributed to the smallness of the time
window; the separation of the time scales, $\tomega \ll \tau \ll 
\twoneeff$, is not enough in the simulations.  

Quite recently Yoshino and the present authors have argued based on the
numerical results on the 4D Ising EA model that fluctuations of
droplets, whose size becomes comparable to that of the preexisting 
domains, become anomalously large, and that they are responsible to the 
occurrence of a peak in $S(t;\tw)$ of the isothermal ZFC magnetization at 
$\tau \simeq \tw$ where $\tau=t-\tw$ is the time elapsed after the 
measuring field is applied.~\cite{YHT} At the end of the transient regime 
of the $T$-shift process, or at the merging to a $T_2$-isothermal aging 
state, similar large fluctuations and so a peak in $S(t; \twone)$ at 
$\tau \simeq \twoneeff$ are expected to appear so long as $\Delta T$ is 
relatively small. In fact this has been experimentally 
observed~\cite{JYN} as already mentioned above. 

Combining the arguments based on our numerical results, in particular, 
the droplets-in-domain scenario and the cumulative memory one, with a 
possible existence of the `chaos effect', we can think of the following 
behavior that a spin glass exhibits in the negative $T$-shift protocol 
of aging depending on the magnitude of $\Delta T$ (and similar behavior 
also for the positive $T$-shift protocol). For a sufficiently small 
$\Delta T$, the merging to the $T_2$-isothermal aging state is observed 
at $\tau \simeq \twoneeff$, where $\twoneeff$ is given by 
Eq.(\ref{eq:tweff}), i.e., by the cumulative memory scenario. 
When  $\Delta T$ becomes large, both the strong rejuvenation just after
the $T$-shift and the merging to the $T_2$-isothermal aging state 
at $\tau \simeq \twoneeff$ are expected to be observed. The latter,
however, becomes to be hardly detected by such an ac susceptibility
analysis done in the present work since $L_{T_2}(\tomega)$ is much
smaller than $L_{T_2}(\twoneeff)$. Instead, it is observed through
a peak in $S(t; \twone)$ of the ZFC magnetization.~\cite{Petra}  
The extracted value of $\twoneeff$ in this case either satisfies the 
cumulative memory scenario ($t_{\rm ov1} \gg \twoneeff$) or is already 
strongly affected by the chaos effect ($t_{\rm ov1} \sim \twoneeff$). 

What is the expected behavior for $T$-shift processes with a sufficiently
large $\Delta T$ for which $t_{\rm ov1} \ll \twoneeff$ or even 
$t_{\rm ov1} \ll t_{\rm min}^{\rm obs}$ holds? Here 
$t_{\rm min}^{\rm obs}$ is the shortest time that the temperature $T_2$
is experimentally stabilized after the $T$-shift. The recent
experimental results are claimed to reach this regime,~\cite{JYN,Petra}
and a theory for the chaos effect on $T$-shift and $T$-cycling processes 
in this regime has been proposed by Yoshino {\it et al.}~\cite{YLB} 
Unfortunately this regime has not been realized in the numerical 
simulations on the 3D EA model. Probably it needs a sufficiently large 
$\twone$, larger than the time-window of our simulation ($\nle 10^5t_0$), 
to realize the condition $R_{T_1}(\twone) > \lovlp$. In order to 
further explore aging dynamics in the $T$-shift protocol, one has to 
systematically choose values of the parameters $\Delta T$ and 
$t_{{\rm w}i}$ even in experiments, since their time-window is similarly 
small ($\sim$ 5 decades) to that of the numerical simulation though its 
absolute magnitude is large ($1s \sim 10^{12}t_0$).

To conclude, we have numerically studied the $T$-shift protocol of aging 
in the 3D EA spin-glass model through the measurement of the ac 
susceptibility. For processes with a small magnitude of the $T$-shift, 
$\Delta T$, the memory imprinted in the first stage of isothermal aging 
is preserved under the $T$-shift and the SG short-range order continuously 
grows with a rate intrinsic to the temperature changed (cumulative memory 
scenario). For $T$-shift processes with a large $\Delta T$ the 
deviation from the cumulative memory scenario has been observed for the 
first time in the numerical simulation. We attribute the phenomenon to 
the `chaos effect' which, we argue, is qualitatively different from the 
so-called the rejuvenation effect observed just after the $T$-shift.

\section*{Acknowledgements}
We thank H. Yoshino for many fruitful discussions, and  P.E. J\"onsson, 
P. Nordblad, V. Dupuis, E. Vincent and H. Mamiya for discussions on their 
experimental results. This work is 
supported by a Grant-in-Aid for Scientific Research Program (\# 12640369), 
and that for the Encouragement of Young Scientists(\# 13740233) from the 
Ministry of Education, Science, Sports, Culture and Technology of Japan. 
The present simulations have been performed using the facilities at the
Supercomputer Center, Institute for Solid State Physics, the University
of Tokyo.


\begin{thebibliography}{99}

%1
\bibitem{rev1}
E. Vincent, J. Hammann, M. Ocio, J.-P. Bouchaud and L.F. Cugliandolo: 
in {\it Proceeding of the Sitges Conference on Glassy Systems},
Ed.: E. Rubi (Springer, Berlin, 1996)

\bibitem{rev2}
J. P. Bouchau, L. F. Cugliandolo, J. Kurchan
and M. M\'ezard: in 
{\it Spin glasses and random fields}, edited by A. P. Young,
(World Scientific, Singapore, 1997).

\bibitem{rev3}
 P. Nordblad and P. Svendlidh: in the same book as Ref. 2. 

\bibitem{JVHBN}
 K. Jonason, E. Vincent, J. Hammann, J.P. Bouchaud, and P. Nordblad:
Phys. Rev. Lett. {\bf 81} (1998) 3243.

\bibitem{revBouch}
 J. P. Bouchaud: cond-mat/9910387.

%6
\bibitem{rejuvenat1}
 F. Lefloch, J.M. Hammann, M. Ocio and E. Vincent:
Europhys. Lett. {\bf 18} (1992) 647.

\bibitem{o-glass}
 P. Doussineau, T. de Lacerda-Ar\^oso and A. Levelut:
 Europhys. Lett. {\bf 46} (1999) 401.

\bibitem{polymer}
 L. Bellon, S. Ciliberto and L. Laroche:
 Europhys. Lett. {\bf 51} (2000) 551.

\bibitem{nano-p}
 P.E. J\"onsson, M.F. Hansen and P. Nordblad:
Phys. Rev. B  {\bf 61} (2000) 1261.

\bibitem{BM-chaos}
A.J. Bray and M.A. Moore:
Phys. Rev. Lett. {\bf 58} (1987) 57.

%11
\bibitem{FH-88-EQ}
 D.S. Fisher and D.A. Huse:
Phys. Rev. B  {\bf 38} (1988) 386.

\bibitem{FH-88-NE}
 D.S. Fisher and D.A. Huse:
Phys. Rev. B  {\bf 38} (1988) 373.

\bibitem{ours1}
 T. Komori, H. Yoshino and H. Takayama:
J. Phys. Soc. Jpn. {\bf 68} (1999) 3387.

\bibitem{ours2}
 T. Komori, H. Yoshino and H. Takayama:
J. Phys. Soc. Jpn. {\bf 69} (2000) 1192.

\bibitem{ours4}
 T. Komori, H. Yoshino and H. Takayama:
J. Phys. Soc. Jpn. {\bf 69} Suppl. A (2000) 355.

%16
\bibitem{Lorenzo-00} 
 L.W. Bernardi, H. Yoshino, K. Hukushima, H. Takayama, A. Tobo and 
A. Ito: Phys. Rev. Lett. {\bf 86} (2001) 720.

\bibitem{Ito-00} 
 A. Ito, A. Tobo, N. Onchi and J. Satooka:
J. Phys. Soc. Jpn. {\bf 69} Suppl. A (2000) 223.

\bibitem{Tc3DJ}
 P.O. Mari and I.A. Campbell: 
 Phys. Rev. E {\bf 59} (1999) 2653.
 
\bibitem{JYN}
P.E. J\"onsson, H. Yoshino and P. Nordblad:
cond-mat/0203444.

\bibitem{PRR}
 M. Picco, F. Ricci-Tersenghi and F. Ritort:
Phys. Rev. B {\bf 63} (2001) 174412.

%21
\bibitem{BB}
 L. Berthier and J.-P. Bouchaud:
cond-mat/0202069

\bibitem{Kisker-96}
 J. Kisker, L. Santen, M. Schreckenberg and H. Rieger:
Phys. Rev. B  {\bf 53} (1996) 6418.

\bibitem{Marinari-98-VFDT}
 E. Marinari, G. Parisi, F. Ricci-Tersenghi and J.J. Ruiz-Lorenzo:
J. Phys. A  {\bf 31} (1998) 2611.

\bibitem{YHT}
 H. Yoshino, K. Hukushima and H. Takayama:
 cond-mat/0202110, 0203267.

\bibitem{JYNKI}
P.E. J\"onsson, H. Yoshino, P. Nordblad, H. Aruga Katori and A. Ito:
cond-mat/0112389.

%26
\bibitem{AlbaHOR-87}
 M. Alba, J. Hammann, M. Ocio and Ph. Refregier:
J. Appl. Phys. {\bf 61} (1987) 3683.

\bibitem{RefregierOHV-88}
 Ph. Refregier, M. Ocio, J. Hammannn and E. Vincent:
 J. Appl. Phys. {\bf 63} (1988) 4343.

\bibitem{Mamiya}
 H. Mamiya, I. Nakatani and T. Furubayashi: 
Phys. Rev. Lett. {\bf 82} (1999) 4332.

\bibitem{VD}
 V. Dupuis and E. Vincent: private communication.

\bibitem{Mamiya-pr}
 H. Mamiya: private communication.

%31
\bibitem{DVBHIK}
 V. Dupuis, E. Vencent, J.-P. Bouchaud, J. Hammann, A. Ito and H. Arga Katori:
 Phys. Rev. B {\bf 64} (2001) 174204.

\bibitem{Kovacs}
 A.J. Kovacs: Adv. Poly. Sci. {\bf 3} (1963) 394; A.J. Kovacs {\it et al}: 
 J. Poly. Sci. {\bf 17} (1979) 1097. 

\bibitem{Petra}
 P.E. J\"onsson, H. Yoshino and P. Nordblad: private communication.

\bibitem{YLB}
 H. Yoshino, A. Lema\^itre and J.-P. Bouchaud:
 Eur. Phys. J. {\bf 20} (2001) 174204.

\end{thebibliography}
\end{document}